\documentclass[manuscript]{aastex}

\usepackage{natbib}
\bibpunct{(}{)}{;}{a}{}{,}

\shorttitle{Discovery of X-ray/UV flare from LMC 335}
\shortauthors{Tsang et al.}

\begin{document}


\title{The Discovery of an X-ray/UV Stellar Flare from the Late-K/Early-M dwarf LMC 335 \\
    }


\author{B. T. H. Tsang\altaffilmark{1}}

\author{C. S. J. Pun\altaffilmark{1}}

\author{R. Di Stefano\altaffilmark{2}}

\author{K. L. Li\altaffilmark{3}}

\author{A. K. H. Kong\altaffilmark{1,3}}


\altaffiltext{1}{Department of Physics, The University of Hong Kong, Pokfulam Road, Hong Kong}
\altaffiltext{2}{Harvard-Smithsonian Center for Astrophysics, 60 Garden Street, Cambridge, MA 02138, USA}
\altaffiltext{3}{Institute of Astronomy and Department of Physics, National Tsing Hua University, Hsinchu, Taiwan}


\begin{abstract}
We report the discovery of an X-ray/UV stellar flare from the source LMC 335, captured by \emph{XMM-Newton}
in the field of the Large Magellanic Cloud. 
The flare event was recorded continuously in X-ray for its first 10 hours from the precursor to the late decay phases.
The observed fluxes increased by more than two orders of magnitude at its peak in X-ray and at least one in the UV as compared to quiescence.
The peak 0.1$-$7.0\,keV X-ray flux is derived from the two-temperature APEC model to be $\sim(8.4 \pm 0.6) \times 10^{-12}$\,erg\,cm$^{-2}$\,s$^{-1}$.
Combining astrometric information from multiple X-ray observations in the quiescent and flare states, we identify the NIR
counterpart of LMC 335 as the 2MASS source J05414534$-$6921512.
The NIR color relations and spectroscopic parallax characterize the source as a Galactic K7$-$M4 dwarf at a foreground distance of (100 $-$ 264)\,pc, 
implying a total energy output of the entire event of $\sim(0.4 - 2.9) \times 10^{35}$\,erg.
This report comprises detailed analyses of this late K / early M dwarf flare event that has the longest time coverage 
yet reported in the literature.
The flare decay can be modeled with two exponential components with timescales of $\sim$28 min and $\sim$4 hours, with a single component decay firmly ruled out.
The X-ray spectra during flare can be described by two components, a dominant high temperature component of $\sim$40$-$60\,MK and
a low temperature component of $\sim$10\,MK, with a flare loop length of about 1.1$-$1.3 stellar radius.
\end{abstract}




\newcommand{\mysrc} {LMC 335}


\section{Introduction}

Stellar flares from magetically active stars provide useful tools to diagnose the physical processes governing the observed multi-wavelength 
characteristics, along with the physical and chemical evolutions of stellar coronae \citep{Benz10}.
Stars of spectral types F, G, K, and M are found to exhibit flare activities of various forms and scales.
Dwarf flares have been extensively studied at wavelengths 
spanning from radio to X-ray, with UV and X-ray observations particularly effective 
for investigations of the cool dwarfs due to their characteristic ranges of coronal temperatures during flare \citep{Osten06, Berger10}. 
Many studies of flare stars have focused on the derivation of information about the magnetic activities associated with spatially unresolved flare events. 
The immense bursts of energy from stellar flares are believed to be generated from dynamo
processes persistent in the stellar interiors and atmospheres. 
Hydrodynamic simulations \citep{Reale97, Favata01, Allred06} and the quasi-static radiative cooling models \citep{vandenOord89, Tsuboi00}
had been widely adopted to study the stellar magnetic activities based on their flare profiles.

Following the early detections of a small sample of flare candidates such as AT Mic and AD Leo \citep{HEAO1det} with the HEAO-1 satellite, 
the \emph{Einstein} and \emph{EXOSAT} observatories extended the study to cover X-ray emission from stellar flares with better sensitivity and longer 
time coverage \citep{EXOSAT90}. 
The observed energetic X-ray flares were divided into two subclasses based on the timescales of the flaring profiles: 
`impulsive' flares with decay times of up to tens of minutes are believed to originate from relatively compact loops similar to those found on the sun;
and `long-decay' flares with typical hours-long decay time observed in solar two-ribbon flares \citep{Priest02}. 
\citet{EXOSAT90} observed 25 K5 to M7 stars within 25\,pc with quiescent (0.05$-$2\,keV) luminosities $\log L$ (erg\,s$^{-1}$) of $\sim$27 $-$ $\sim$30,
and found X-ray flares with energies from $\sim$3 $\times 10^{30}$ to $\rm 1 \times 10^{34}\ erg$, and
flare temperatures of $\sim$20$-$40\,MK. These values are comparable to those obtained from \emph{Einstein} and these observations
have extended the studies on stellar flares with improved sensitivity especially for their quiescent emissions.

In the extreme UV (EUV) wavelength, \citet{CompLoops} conducted a survey on 44 type F to M stars in the Extreme UltraViolet Explorer 
(EUVE) database containing a total of 134 flares to explore the underlying dynamo mechanisms using physical parameters of the 
flare loops derived from a homogeneous set of flare lightcurves. 
The primary parameters investigated are the flare loop length and the magnetic fields during flares.
It was suggested that the observed increase of maximum flare loop sizes from the warmer K2 to M0 sample to the cooler and later ones could be due to
a transition from a core-convection dynamo to a turbulent dynamo.

After scattered \emph{Einstein} and \emph{EXOSAT} observations, the R\"{o}ntgen Satellite (\emph{ROSAT}) All-Sky Survey (RASS) provided 
a comprehensive coverage for the entire sky in soft X-ray. 
\citet{Tsikoudi00} selected a sample of 14 active stars from K0 to M5.5 in the RASS database to search for
concurrent energetic flares in both EUV and X-ray energies. 
Due to the observing strategy in the scanning mode, all the detected flares are of longer duration with event ages longer than an hour.
High correlations of the X-ray (0.1$-$2.4\,keV) and EUV (60$-$140\,\AA\ or 110$-$200\,\AA) luminosities could be deduced for all observed flares,
with flare luminosities $\log L$ (erg\,s$^{-1}$) from 27.3 $-$ 30.0 for X-ray, and 26.6 $-$ 29.7 in EUV. 

The better spatial and spectral resolution of the \emph{XMM-Newton} and \emph{Chandra} X-ray observatories vastly enhance the investigations of flare 
stars especially in the lower temperature end. Flare emissions from a number of dwarfs later than M8 were
observed \citep{Rutledge00, Stelzer06, Robrade09, Robrade10, Gupta11}. 
These observations also enabled quiescent emission of late type stars to be detected and studied at much fainter limits, 
revealing the highly unstable nature during quiescence, commonly referred to as quasi-quiescence states. 
Simultaneous monitoring in the X-ray and optical/UV wavelengths by \emph{XMM-Newton} provide further constraints on the possible emission mechanisms
associated with the magnetic activities. 
The availability of high-resolution X-ray spectroscopy supports the studies of coronal elemental abundances \citep{Audard03, Sanz-Forcada04} 
as well as the construction of emission measure distribution during different stages of the flare evolution for detailed investigations of the 
temperature structure of the stellar coronae \citep{Huenemoerder01, Raassen03}.

The discovery of a previously unclassified flare object in the field of LMC by \emph{XMM-Newton}, 
first detected in the quiescent state by the \emph{ROSAT} High Resolution Imager (HRI) survey \citep{Sasaki00}, is reported in this paper. 
The detected X-ray flare is found to be positioned at $\sim$2.5\arcsec\ from the catalogued position of 
source 335 \citep{Sasaki00}, hereafter referred to as LMC 335.
In addition to \emph{ROSAT} and \emph{XMM-Newton}, the source was also observed in X-ray quiescence with the \emph{Chandra} and \emph{Swift} satellites.
In Section \ref{Obs} the X-ray flare and quiescent observations of \mysrc, along with UV and optical images of the field, are described. Results of 
the findings are presented in Section \ref{Results}: with Section \ref{SrcPosID} focusing on the source and counterpart identification, 
and Sections \ref{Timing} and \ref{Spectral} on the timing and spectral analysis respectively. 
Discussions and conclusions are presented in Section \ref{Discussions}.

--------------------

\section{Observations and data analysis}
\label{Obs}

After the first detection by the \emph{ROSAT HRI}, \mysrc\ had been captured in X-ray  
with various instruments including \emph{XMM-Newton}, 
\emph{Chandra}, and \emph{Swift}. Only
one of the \emph{XMM-Newton} observations captured an X-ray/UV flare event, while all the other observations
were made during quiescent states. Observational data in X-ray were
obtained from the High Energy Astrophysics Science Archive Research Center (HEASARC) database.
In this section we summarize the multiwavelength observations of \mysrc\ and highlight those related to the flare event. 

In Table \ref{xmmsumtable} we list all the \emph{XMM-Newton} European Photon Imaging Camera (EPIC) observations that included \mysrc. 
These observations were originally targeted at the pulsar PSR0540$-$69 and thus \mysrc\ was at an off-axis angle of $\sim$9\arcmin.
The analysis was performed using the standard \emph{XMM-Newton} Science Analysis System (SAS) version 11.0.0. 
With a majority of the quiescent state and flaring state X-ray photons contained in the lower energy 
bins, the energy band used for exposure map corrections and source detection was limited to 0.1 $-$ 7.0\,keV.
At the position of \mysrc, source spectra were extracted with the SAS task \emph{evselect} using a circular region
of radius 10\arcsec, while the background was measured from a nearby source free circular region of radius 40\arcsec\ on the same CCD. 
The same regions were adopted to generate lightcurves, with the task \emph{epiclccoor} used to perform the background subtraction.
All the spectra and response files were grouped with minimum counts of 20 and 10 for the flare duration and the quiescent phase respectively. The grouped spectra were then analyzed with XSPEC version 12.7.0.

The X-ray flare event from \mysrc\ was recorded by the two \emph{XMM-Newton} EPIC MOS cameras in medium
optical filters on 2000 May 27 during a 40\,ks observation (Obs.ID 0125120201 in Table \ref{xmmsumtable}).
During the observation, while the central CCDs (CCD 1) of the two MOS cameras were operating in the Timing mode,
the outer CCDs (CCD 2 $-$ 7), where \mysrc\ was captured, were in the Imaging mode. Unfortunately, the more sensitive EPIC PN camera was
in the Small Window mode and did not include \mysrc\ in its field of view. Therefore only data from two MOS cameras are
available for the X-ray flare analysis. 

In addition to the EPIC cameras, the Optical Monitor (OM) onboard \emph{XMM-Newton} also recorded simultaneous UV broadband 
observations of \mysrc. 
Three of the ten \emph{XMM-Newton} observations have OM imaging coverage, one of which was fortuituously taken during the flare.
During the flaring observation, the OM took a series of UV images with filters UVW1 (245 $-$\,320 nm), UVM2 (205 $-$ 245\,nm), and UVW2 (180 $-$ 225\,nm), 
with exposure time ranging from 1\,ks to 4\,ks. The exposures were all taken with the lateral Window 1 with 2 $ \times $ 2 pixel binning.
Details of the OM observations are summarized in Table \ref{OMUVTab}. 
\mysrc\ was not detected by the OM in the UV except for one UVW1 observation taken roughly 0.5 hours after the X-ray flux maximum. 
For the non-detections, the 3-$\rm \sigma$ upper limits of count rates, magnitudes, and fluxes, were estimated with the average count rate
computed from five source-free background regions near the source position. 
The detected magnitude was taken from the \emph{omichain}-generated Observation Window (OSW) source lists. 
The magnitude to UV flux and upper limits conversion were done using the standard reference fluxes of Vega. 

X-ray emission from the quiescent phase of \mysrc\ was independently detected by the \emph{Chandra} and \emph{Swift} satellites in addition
to \emph{XMM-Newton}. 
In the case of \emph{Chandra}, three HRC-I Non-grating observations of \mysrc\ were taken in three separate pointings on 1999 August 31 (Obs.ID 132) and
2000 June 21 (Obs.IDs 1735 and 1736). The event lists were merged using the Chandra Interactive Analysis of Observations (CIAO)
version 4.2 task \emph{dmmerge} with an effective exposure time of $ \sim $48\,ks. 
The observed quiescent count rate of (3.4 $\pm$ 0.5) $\times 10^{-3}$\,s$^{-1}$ was consistent with that measured by the \emph{XMM-Newton}.
On the other hand, the better angular resolution of the HRC data ($<$0.5\arcsec) provides more accurate position than 
the \emph{XMM-Newton} MOS data (2.0\arcsec). 

The field of \mysrc\ had been imaged by \emph{Swift} in a series of observations targeted also at PSR0540$-$69. A total of 22 sets of data were obtained
spanning from 2005 April to 2010 November with an integrated exposure time of $\sim$170\,ks. Similar to the \emph{Chandra} observations, the data were 
taken when \mysrc\ was in quiescent states and the X-ray emission was weak, with count rate of (1.1 $\pm$ 0.2) $\times 10^{-3}$\,counts\,$^{-1}$. 
The event lists were combined before lightcurves and spectra were extracted with the \emph{Swift} tool \emph{xrtgrblcspec}. 
A low signal-to-noise spectrum could be extracted from the combined event list, and was found to be consistent with (but of poorer quality than) the \emph{XMM-Newton} data. 
The X-ray position and positional errors were computed using the task \emph{xrtcentroid} for comparison with the other observations.

In addition to X-ray observations, archival optical/NIR data taken with European Southern Observatory Wide Field Imager (ESO-WFI) of \mysrc\ 
are available through the observations of the association LH 104. 
The raw ESO-WFI images taken on 2004 November 21 were obtained from the ESO Archive\footnote{http://archive.eso.org/eso/eso\_archive\_main.html}
in four filters, \emph{B} Bessel (ESO\# 878), \emph{Rc}/ 162 (ESO\# 844), Z+60 (ESO\# 846), and H\_alpha 7 (ESO\# 856),
with cumulative exposure time of about 4, 2, 43, and 25 minutes respectively. 
With the reduced images, astrometry was done by registering five 2MASS and one Tycho bright stars near the field of \mysrc\ with no known proper motions. 
The errors of the WCS mapping for all the ESO-WFI images are well below 0.1\arcsec\ near the source \mysrc. 
Two catalogued sources, 2MASS J05414534$-$6921512 \citep{tMASS} and DENIS J054145.2$-$692151 \citep{DENIS}, are found to be closest to the
position of \mysrc\ and are considered to be the IR counterparts of the source. Since there is no systematic differences in $J$ and $K_{S}$ band
measurements between 2MASS and DENIS \citep{Cabrera03}, the consistent $J$ and $K_{S}$ band magnitudes together with the slight positional offset 
of $\sim$0.4\arcsec\ of these two catalogued sources suggest that these counterparts are indeed the same physical object. 

\section{Results}
\label{Results}

\subsection{Source Position and Near-IR counterpart}
\label{SrcPosID}
\begin{figure}
\epsscale{.80}
\fbox{
\plotone{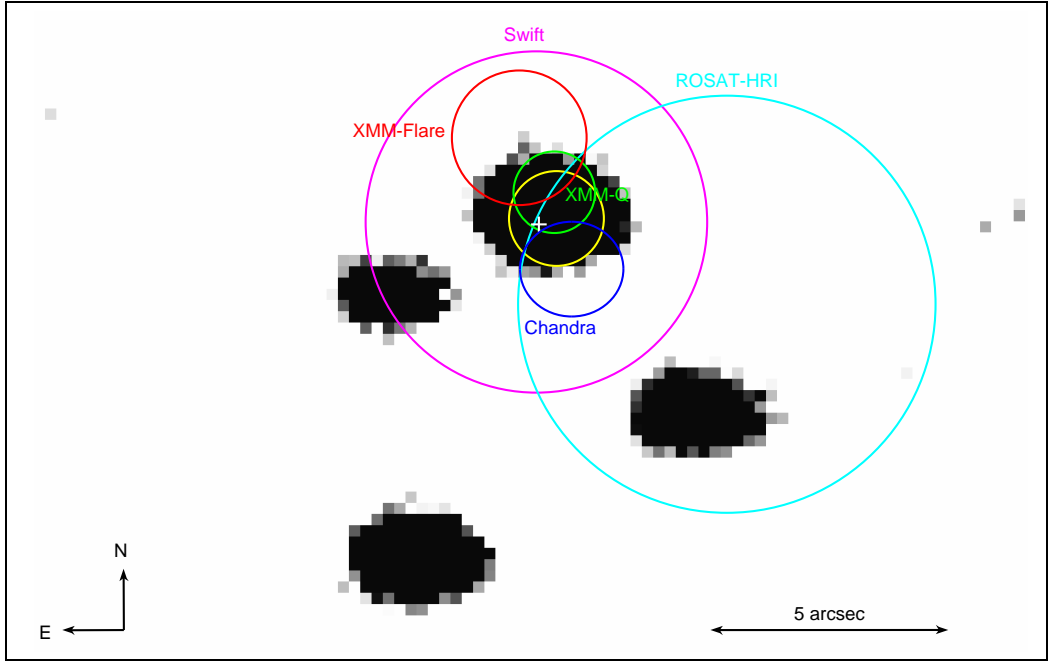}
}
\caption{ESO WFI $R_{c}$ band image of the \mysrc\ field shown with NIR and X-ray source positions and 1-$\sigma$ error circles.
White cross: 2MASS J05414534$-$6921512 counterpart, yellow circle: DENIS J054145.2$-$692151 counterpart, 
red circle: weighted average position from \emph{XMM-Newton} flare observations, 
green circle: weighted average position from the \emph{XMM-Newton} quiescent observations,
blue circle: \emph{Chandra} HRC position, 
cyan circle: \emph{ROSAT} HRI position,
magneta circle: \emph{Swift} XRT positon.
\label{wfiRsrcpos}}
\end{figure}

Identification of counterparts of \mysrc\ could reveal the nature of the X-ray flare event. 
Positions of the source computed from all the X-ray and IR data are listed with the 1-$\rm \sigma$ errors in Table \ref{srcpostable}. 
For the \emph{Chandra} HRC data, the reported position for the source at quiescent state was obtained by the result obtained from image binning factors of 16. 
In the case of \emph{XMM-Newton}, both the flare and quiescent state positions are reported. The two MOS flare detections were used to determine the 
flare state position while all other detections were used for quiescent state measurement. 
Before weight averaging the results from the individual detections, the 1-$\rm \sigma$ positional errors were obtained by
combining in quadrature the statistical maximum likelihood (ML) fitting errors with the systematic error of the pointing of \emph{XMM-Newton}, 
taken to be 2.0\arcsec\ \citep{Guainazzi11} in this work.
A similar approach of positional error calculation was adopted for the \emph{XMM-Newton} Serendipitous source catalogue \citep{tXMM}. 
With the best angular resolution, the \emph{Chandra} HRC data provides
the smallest uncertainty in the X-ray position of the source at its quiescent state, despite the low count rate recorded for the data. 
On the contrary, the low resolution of \emph{Swift} and \emph{ROSAT} data imply poor astrometry. 
The 1-$\rm \sigma $ uncertainty contours of the quiescent and flare states X-ray emission are overlaid on the ESO-WFI \emph{Rc}-band image of the field 
in Figure \ref{wfiRsrcpos}, along with the catalogued positions of the two infrared counterparts 2MASS J05414534$-$6921512 and DENIS J054145.2$-$692151.
The 2MASS position errors are too small to be presented as an error ellipse and its position is thus represented by a cross.
It can be seen from Figure \ref{wfiRsrcpos} that all X-ray detections are consistent in position with the uncatalogued ESO-WFI source that has 
IR counterparts from both the 2MASS and DENIS catalogues, with only slight deviation for the flare state detection position at 
$\sim$2\arcsec\ from the IR source positions. Offsets of similar extent were found with the other X-ray bright source CAL 69 in the 
same \emph{XMM-Newton} field, suggesting that the flare state position is compatible with the NIR counterparts identified.

In order to identify the nature of the source \mysrc, we studied the magnitudes and colors of the IR counterparts.
The results are listed in Table \ref{counterpartinfo}. From its colors and magnitudes, it is possible to estimate the spectral type of \mysrc. 
Comparing our results with the 2MASS photometry of a filtered set of 461 K5 to M9.5 Galactic stars \citep{DwarfDatabase}, we found 90 K7 to M5.5 stars
having 2MASS color indices lying within the error range reported in Table \ref{counterpartinfo}. The sources are mainly in the early M types,
and has a median type of M2.5, with 95.6\% (86/90) lying between K7 to M4.
This is consistent with the Galactic K to M dwarfs identification we derived from the stellar population study along the direction to 
the LMC by \citet{stellarpoplmctmass} using the 2MASS $J - K_{S}$ and $K_{S}$ color-magnitude diagram. 
Using the spectroscopic parallax relation based on a sample of MLT dwarfs by \citet{MLTSDSS}, and assuming a spectral type of K7 to M4, 
the absolute M$_{J}$ magnitude of the 2MASS counterpart is
estimated to be 6.20$-$8.34, which transforms into a distance of 100$-$264\,pc, with the median distance of 174\,pc.
Such estimation suggests the source is not in the LMC but is in the 
foreground. This distance range will be adopted as the distance to the object in the subsequent analyses of X-ray/UV emission.

\subsection{Timing Analysis}
\label{Timing}
\begin{figure}
\epsscale{.80}
\centering
\includegraphics[scale=0.6, angle=270]{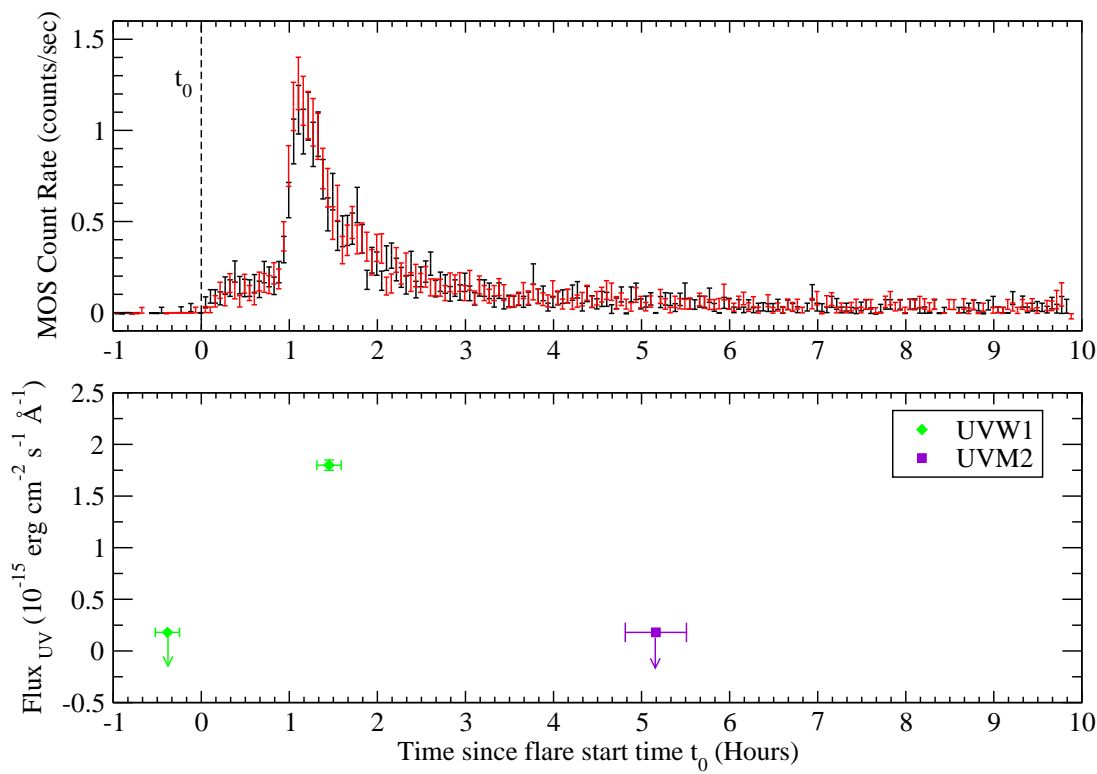}
\caption{
\emph{Upper panel:} \emph{XMM-Newton} EPIC MOS background subtracted lightcurves (black: MOS1, red: MOS2) of \mysrc\ in 0.1$-$7.0\,keV band
throughout the flare event. 
Time t = 0 corresponds to 2000 May 27 UT 02:50, the flare start time denoted by t$_{0}$. 
Time bins used are 200\,s.
\emph{Lower panel:} The \emph{XMM-Newton} OM-derived UV fluxes/upper limits in the corresponding filters during the X-ray observations.
\label{MOSlc}}
\end{figure}

The X-ray flare event of \mysrc\ on 2000 May 27 was captured by the \emph{XMM-Newton} MOS cameras in its entirety. 
The 0.1$-$7.0\,keV background-subtracted X-ray lightcurve 
with 200\,s binning is presented in the upper panel of
Figure \ref{MOSlc} (black: MOS1; red: MOS2). The time axis starts at 2000 May 27 UT 02:50 when the precursor phase (P) started, this will be referred to as
the \emph{flare start time t$_{0}$} from hereafter.
The quiescent state (Q) of the source persisted during the first hour of observations with flux beneath our detection limit. Only a 3$\rm \sigma$ 
upper limit count rate of about 6 $\times$ $10^{-3}$\,$\rm s^{-1} $ could be obtained. This is consistent with the expected average quiescent count rate 
computed from all other \emph{XMM-Newton} MOS detections at (4.0 $\pm$ 0.7) $\times$ $10^{-3}$\,$\rm s^{-1} $. 
The \emph{Chandra} and \emph{Swift} quiescent observations also produced 
consistent values, with PIMMS-converted (v4.4) \emph{XMM-Newton} MOS count rates being (5.7 $\pm$ 0.9) $\times$ $10^{-3}$ 
and (3.9 $\pm$ 0.6) $\times$ $10^{-3}$ respectively, assuming a 1.1\,keV ($\sim$12\,MK) APEC plasma model (c.f. Table \ref{SPECFitTab}, 
Section \ref{Spectral}).
At $\sim$1 hour after the start of the observation, the X-ray count rate suddenly rose to a precursor level (P) of 0.1 $-$ 0.2\,$\rm s^{-1}$. 
At $\sim$55 min after $t_{0}$, the flare entered the rise (R) phase when 
the count rate rose sharply during $\sim$800\,s ($\sim$13 min) from the precusor level to over 
1.1\,$\rm s^{-1} $ at the flare peak. 
The flare flux then entered a decay phase starting 1.13 hours after $t_{0}$ and continuing until the end of the observation almost 10 hours later.

The OM provides simultaneous UV observations during the X-ray flare of \mysrc. Flux variation in the UV band during the flare is shown in 
Figure \ref{MOSlc} (lower panel).
Detailed detected count rate, magnitude, and flux, as well as the 3-$\rm \sigma$ upper limits, are listed in Table \ref{OMUVTab}. 
Only upper limits could be derived for the UVW1 observation taken during the quiescent phase of the flare, and the UVM2 data taken $\sim$5 hours after 
$t_{0}$. The only detection was the UVW1 image taken $\sim$1.5 hours after $t_{0}$ during the Decay 1 (D1) phase, when a UV flux rose by at least
one order of magnitude from the quiescent phase upper limit.
\citet{MitraKraev05} studied the relation between X-ray and UV energy budget and found that the luminosity increases in X-ray and UV
(measured in UVW1 band) during a stellar flare can be expressed as 
$\Delta L_{\rm X} \propto \Delta L_{\rm UV}^{\alpha}$, where $\alpha$ $\sim$1.1$-$1.2. 
While the time resolution of our UV data is not high enough for more
detailed analysis, the observed flux increases in X-ray and UV are consistent with the power-law relation.

\begin{figure}
\epsscale{.80}
\plotone{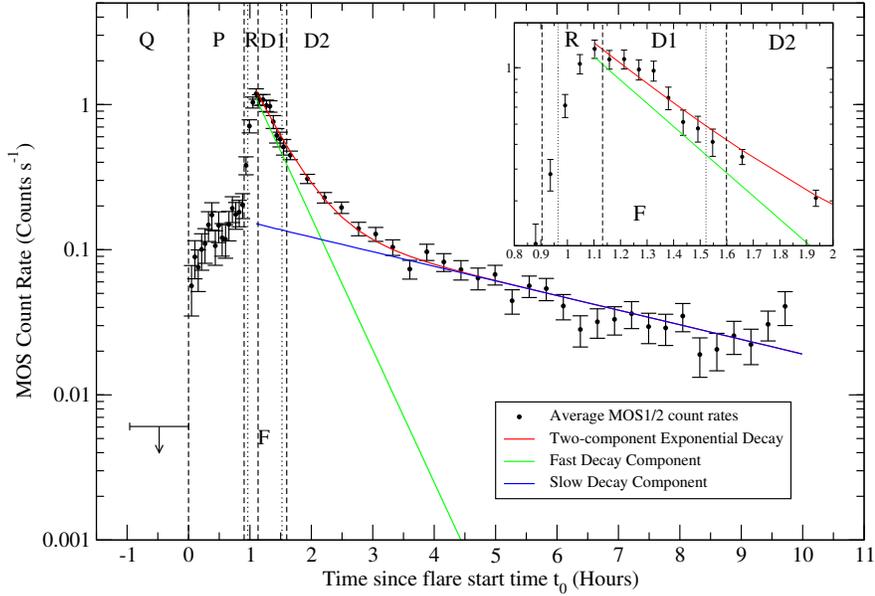}
\centering
\caption{
MOS1 and MOS2 averaged lightcurve of \mysrc\ is replotted in logarithmic scale overlaid with the exponential fits.
Time t = 0 refers to 2000 May 27 UT 02:50, the flare start time t$_{0}$ as denoted in Figure \ref{MOSlc}.
The observation is partitioned into five phases: quiescence (Q), precursor (P),
rise (R), decay 1 and 2 (D1$-$D2) marked by black dashed lines. The additional flare
(F) phase bounded by dotted lines is added to study the properties of \mysrc\ during the peak state. 
The time bins for D2 have been widened from 200\,s to 1000\,s for better statistics and visualization. 
Detailed evaluation of the lightcurve near the flare peak is shown in the inset figure.
\label{logLC}}
\end{figure}

To facilitate a quantitative study of the behavior of the flare during the decay phase, the lightcurve data points taken after $t_{0}$ + 1.6 hours
were rebinned
into intervals of 1000\,s to give better statistics. The MOS1-MOS2 averaged count rates from the peak of
emission 1.13 hours after $t_{0}$ are fitted with various exponential decay functions. 
We found that the data could be fit well with 
a two-component exponential decay function (reduced $\chi ^{2}$, DOF = 1.01, 35) while the single-component model can be ruled out due to apparent deviation 
especially after the first hour of decay (reduced $\chi ^{2}$, DOF = 17.07, 37).
Initially the drop was rapid with an e-folding time (and 1-$\sigma$ error) of 28.6$^{+1.9}_{-1.3}$ min, the decay then slowed down to an e-folding time 
of 4.3$^{+3.5}_{-0.6}$ hours. 
The X-ray lightcurve of the flare and the exponential fit are shown in Figure \ref{logLC}. 

Based on the exponential fits to the X-ray lightcurve, we further divided the decay phase into the Decay 1 (D1) phase from the flare peak
to include the first e-folding time of the fast decay component, followed by the Decay 2 (D2) phase until the end of the observations. 
The time boundaries of the Quiescence (Q), Precursor (P), Rise (R), Decay 1 (D1) and Decay (D2) phases are marked by black vertical dashed lines 
in Figure \ref{logLC}. 
To have a more detailed analysis of the conditions of the flare during its peak, an additional Flare (F) phase overlapping with R and D1, 
starting from 10.0 min before the peak to 23.4 min after flare peak, is defined as the interval during which the detected X-ray count rate is higher than half of the peak value (marked with dotted lines in Figure \ref{logLC}).
The inset in Figure \ref{logLC} shows a zoomed-in view of the lightcurve near the flare peak. At the beginning of phase D1, a slight
excess in counts is observed at about 1.3 hours after $t_{0}$, which could possibly come from another very brief rise phase after the main peak.
Such multiple peak lightcurves had been detected in other stellar flares with comparable rise times \citep{Pandey08, GALEXUV}.
On the other hand, this slight excess could certainly also have been resulted from the statistical fluctuations of the data.

This flare observation of \mysrc\ was fortuitously captured near the beginning of a long \emph{XMM-Newton} exposure, thus allowing continuous observations
of the first 10 hours of the flare event.
From Figure \ref{logLC}, it is obvious that the X-ray emission still had not returned to the quiescent emission level 
by the end of the observation. The average MOS count rate for the last hour before the end of observations is 
0.029 $\pm$ 0.006\,$\rm s^{-1}$, which is a factor of 5 to 7 higher than the expected quiescent rate of 0.004\,$\rm s^{-1}$.
This could originate from low-level flares persistent in the active regions after the main event.

\subsection{X-ray Spectral Analysis}
\label{Spectral}
\begin{figure}
\epsscale{.80}
\plotone{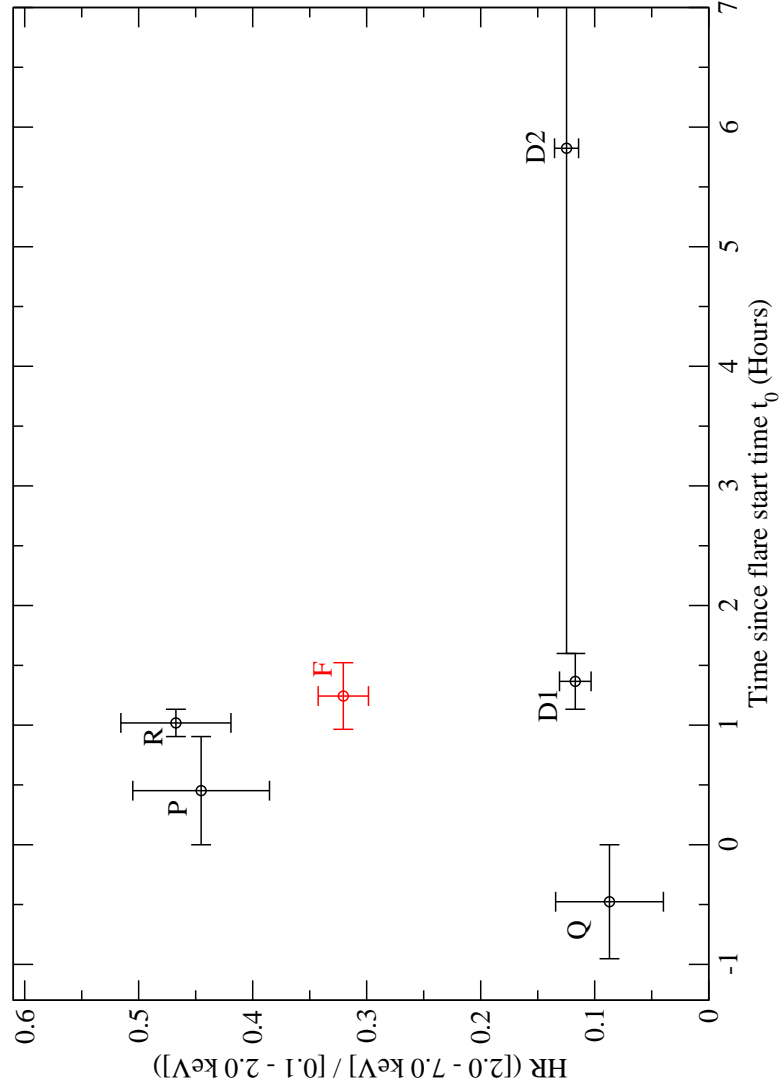}
\centering
\centering
\caption{
Evolution of X-ray Hardness Ratio (HR) during the six representative phases from quiescence (Q) to decay 2 (D2).
The averaged HR from the non-flare PN \emph{XMM-Newton} observations is used to represent phase Q.
\label{HRplot}}
\end{figure}

With the count rate at a low level for most of the duration of the flare, hardness ratios (HRs) were first computed to study the overall X-ray energy
distribution of the event and to provide model-independent information for comparisons with other flares and other X-ray sources.
The flare is soft, with X-ray photons in the energy range 0.1$-$2.0\,keV making up $\sim$80\% of all photons collected during the 
observation. The HR is defined as $\rm HR = H / S $, where S and H are respectively the $\rm 0.1 - 2.0\,keV$ and $\rm 2.0 - 7.0\,keV$
count rates averaged from the MOS1 and MOS2 observations. 
Since \mysrc\ was below the detection limit during the first hour of the flare \emph{XMM-Newton} MOS observations,
the HR for the quiescent phase is averaged from the non-flare PN \emph{XMM-Newton} observations for proper comparison with the spectral fitting results. 
Figure \ref{HRplot} shows the time evolution of the HR of the flare during the 
five phases from pre-flare quiescence (Q) through the decay phases, with the flare phase (F) data point also shown for reference. 
We observed a sharp rise in HR from Q at 0.09 $\pm$ 0.05 to P phase of 0.45 $\pm$ 0.06,
reaching a maximum at phase R of 0.47 $\pm$ 0.05. The value of HR then dropped through the F and the two decay phases, 
settling back near the quiescent level at 0.12 $\pm$ 0.01 for the D2 phase. 
The results imply that the flare in hard X-rays commenced simultaneously with the flare of the softer components, but decayed away much more quickly. 
This observed softening in X-ray after the R phase is common among late-type stellar flares \citep{Reale04, Treholme04, Schmitt02}.

In addition to computing the hardness ratios, we have investigated the X-ray spectra of the flare in each phase. 
Identical procedures were followed for the generation of X-ray spectra in each phase except for the different numbers of minimum counts 
used to bin the spectra. 
The single-temperature (1T) and two-temperature (2T) thermal emission APEC models were used to analyze the spectra, except for the Q phase
spectrum in which only the 1T model was carried out 
to derive a best-fit temperature of 1.22\,keV (14.2\,MK). 
The 2T model was found to provide better fits than the 1T model for all the other phases, with significant improvements for phase D1, D2, and F. 
As the fits were insensitive to the choice of metal abundance parameter, the Solar abundance was therefore used.
The \emph{EM} is derived from the normalization of the APEC model fitting in XSPEC and the assumed source distance range.

\begin{figure}
\epsscale{.80}
\plotone{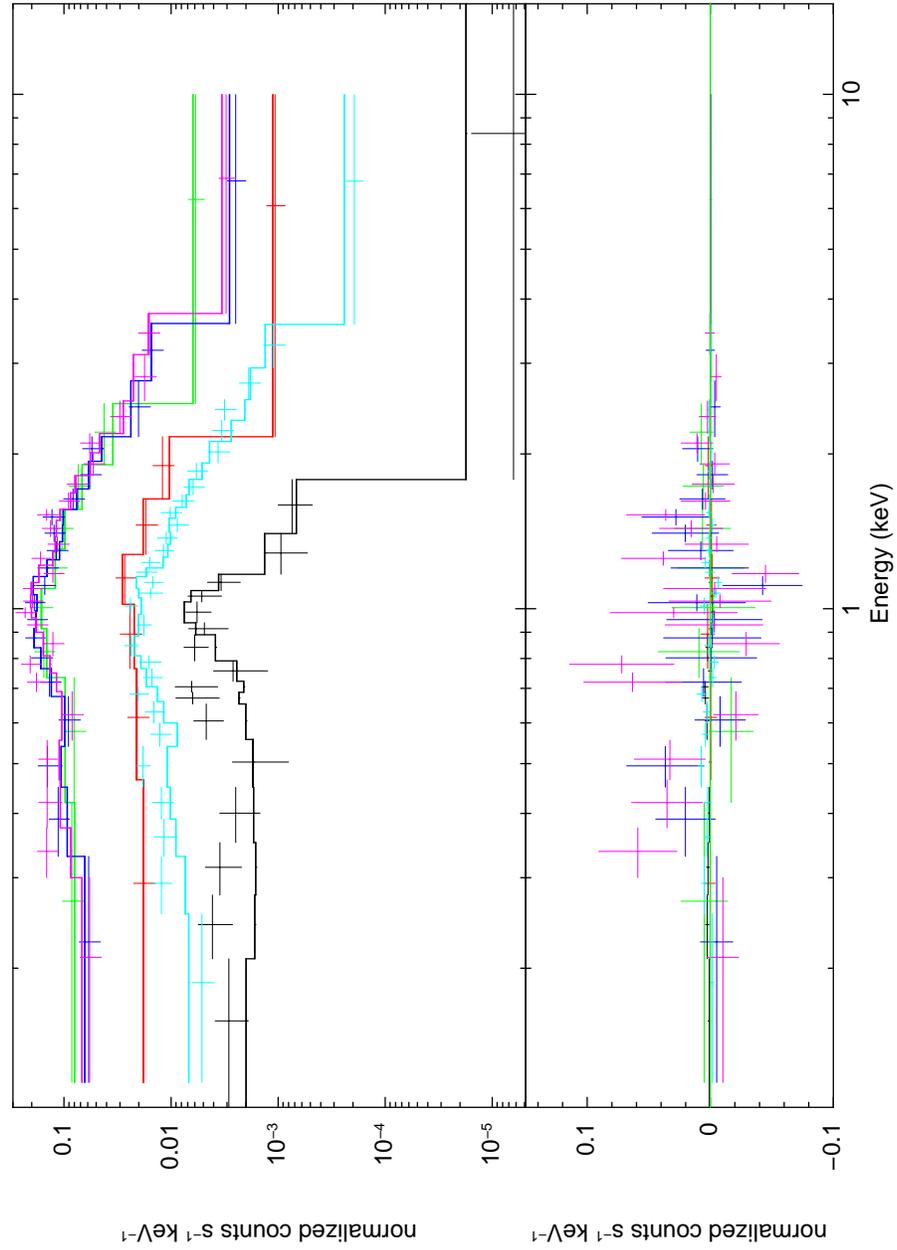}
\caption{\emph{XMM-Newton} MOS1 background subtracted spectra with best fit 1T APEC model for
quiescence (black); and 2T APEC model for precursor (red), rise (green), decay 1 (blue), decay 2 (cyan), and flare (magneta) phases.
\label{M1MOSspec}}
\end{figure}

We observed that the high temperature components of $\sim$5\,keV emerged in the
P and R phases, followed by a slight decline during the decay phases to $\sim$3\,keV while the low temperature components stayed at $\sim$1\,keV.
The dominance of the hotter components is revealed from the consistently larger \emph{EM} throughout the event as well as 
the biases in the fitted temperatures towards the hot range with the 1T model.
Also, we observed an increase in \emph{EM} for the cooler components above the quiescent value particularly in the R, D1, and F phases, 
which could be associated with the flare.
Noticeable increases (by factors of $\sim4 - 9$) in \emph{EM} of the low temperature components from quiescence to flare peaks 
had been observed in other flares \citep{Robrade10, Gupta11}, but not as high as observed for this object (factor of $\sim$23).
Since the spectra from MOS1 and MOS2 are very similar, only the MOS1 spectra taken during each phase are shown in Figure \ref{M1MOSspec}.
Data from the MOS1 and MOS2  were fitted together to provide better statistics 
and the quiescent spectrum is obtained by previous non-flare PN observations.
The best-fit parameters of the two models in different flare phases are summarized in Table \ref{SPECFitTab} with all 
errors given in 90\% confidence intervals as computed in XSPEC.
The spectra had also been fitted by the equilibrium plasma model Mewe-Kaastra-Liedahl (MeKaL), resulting in similar best-fit parameters.

Due to the limited signals in the quiescent (Q) and the precursor (P) phases, the line-of-sight neutral hydrogen column density ($n\rm _{H}$) 
could not be constrained by the fitting. Fixing $n\rm _{H}$ to values below 10$^{21}$\,cm$^{-2}$ did not induce significant changes to the 
fitted plasma temperature, while the fits were poor (with reduced $\chi ^{2} \ge 3$) for $n\rm _{H}$ values above a few times of 10$\rm ^{21}$\,cm$^{-2}$. 
The $n\rm _{H}$ in quiescent and precursor states were thus fixed to be the average of the best-fits values from the two decay phases D1 and D2 at 
0.67 $\times 10^{20}$\,cm$^{-2}$ for both the 1T and 2T models. 
The value from the rise phase (R) was not used for the estimation of $n\rm _{H}$ due to the unstable physical conditions 
during the period and the resulting bigger uncertainties of the fitted parameters.
By using a fixed $n\rm _{H}$ in all phases, the other spectral parameters are similar to those obtained with floating $n\rm _{H}$.
The best-fit values of $n\rm _{H}$ from these spectra were about one order of magnitude lower than the weighted average values of hydrogen column density
along that line-of-sight towards the LMC estimated from the Leiden/Argentine/Bonn (LAB) survey ($\rm 3.37\ \times 10^{21}$\,$\rm cm^{-2}$) \citep{LABSur}
and the Dickey \& Lockman (DL) Galactic survey ($\rm 6.76\ \times 10^{20}$\,$\rm cm^{-2}$) \citep{DLMap}. 
This provides additional supporting argument that the source is indeed located in the foreground instead of within the LMC.

The luminosity of \mysrc\ at quiescent state $L_{\rm X,q} =$ (0.2 $-$ 1.7) $\times 10^{29} $\,erg\,s$^{-1}$ (0.1 $-$ 7.0\,keV)
measured by \emph{XMM-Newton} is comparable 
with the PIMMS implied value from the \emph{ROSAT} HRI detected value of
2.4$\times 10^{29}$\,erg\,s$^{-1}$ (0.1 $-$ 2.4\,keV) assuming the 1.1\,keV APEC model as deduced for quiescence.
\emph{Chandra} and \emph{Swift} observations also support the relatively stable X-ray activity of the source in quiescence, 
with inferred luminosities of 1.8$\times 10^{29} $\,erg\,s$^{-1}$ (0.08 $-$ 10.0\,keV) 
and 1.3$\times 10^{29} $\,erg\,s$^{-1}$ (0.2 $-$ 10.0\,keV)
respectively assuming the same spectral model.

Combining and averaging from the two MOS observations for the flare during the entire duration of observation,
the flux of the source is $(8.8 \pm 0.4) \times 10^{-13}$\,erg\,s$^{-1}$\,cm$^{-2}$, 
implying a total emitted X-ray energy of (0.4 $-$ 2.9) $\times 10^{35}$\,erg  
assuming a distance to the source of (100 $-$ 264)\,pc and the total exposure time of 40\,ks. 

To characterize the X-ray activity, the $\log (L_{\rm X}/L_{\rm bol})$ ratios at different stages were computed and compared. 
To estimate the bolometric luminosities from the bolometric correction (BC) factor, the BC$_K $-($J - K_{S}$) relation derived from 51 M0$-$L7 stars 
with 0.75 $\leq$ ($J - K_{S}$) $\leq$ 1.60 by \citet{BCKdML} was used. To construct the BC relation, they computed the bolometric luminosities of the sample
stars by integrating the observed IR spectra, with the short-wavelength ends linearly extrapolated to zero flux at zero wavelength and the long-wavelength
ends approximated by the Rayleigh-Jeans tail.
Using the IR colors of the 2MASS counterpart in Table \ref{counterpartinfo}, the BC$_K $ was computed to be 2.63, 
and this transforms into an apparent bolometric magnitude of 15.10. 
Combining with the source distance of (100 $-$ 264)\,pc, an absolute bolometric magnitude
$M_{\rm bol}$ of (8.0 $-$ 10.1) could be determined, 
implying the value of $\log(L_{\rm bol}/L_\odot)$ from $-$2.1 to $-$1.3.
The range agrees closely with the estimates of $-1.27$ and $-1.99$ for M0/M1 and M3/M4 stars respectively in the \emph{GALEX} M dwarfs 
UV flare survey \citep{GALEXUV}. 
The APEC model results were used for estimation of the total X-ray luminosity $L_{\rm X}$ to yield 
$\log (L_{\rm X,q}/L_{\rm bol}) =$ $-$3.1 during quiescent state and $\log (L_{\rm X,flare}/L_{\rm bol}) = -0.4$ in the flare state, 
suggesting a significant rise in X-ray activity. 

\begin{figure}
\epsscale{.80}
\plotone{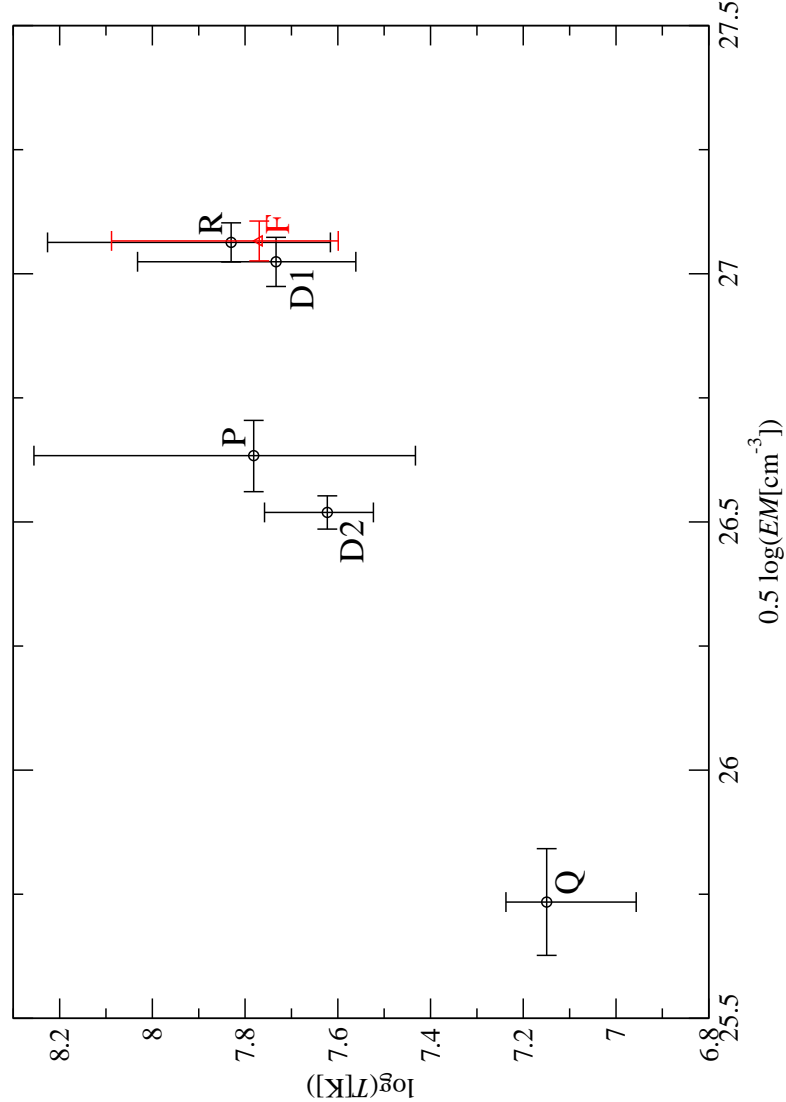}
\centering
\centering
\caption{
Time evolution of temperature and the Emission Measure (\emph{EM}) during the six representative phases from quiescence (Q) to decay 2 (D2) as
derived for a reference spectral type of M2.5 (174\,pc). 
The data point for quiescent phase is derived from the 1T APEC model; parameters for all other phases are obtained from the high-temperature 
components of the 2T APEC fits. The error bars for the \emph{EM} values include only the uncertainties in XSPEC normalization, 
the uncertainty in source distance is not included.
For spectral types of K7 (264\,pc) and M4 (100\,pc), the values on the \emph{EM} axis offset by 0.18 and $-$0.24 respectively with the relative
positions of data points unchanged.
\label{T_EM}}
\end{figure}

The evolution of the flare from quiescent to the decay phase is traced on the $\log T - \log \sqrt{EM} $ plane as shown in Figure \ref{T_EM}. 
All the \emph{EM} values were taken from the dominant high-temperature components of the 2T spectral fits except for phase Q.
Data points on Figure \ref{T_EM} are calculated based on the reference spectral type of M2.5 (174\,pc),
and the \emph{EM} error bars includes only spectral fitting errors with XSPEC, while errors due to the uncertain source distance are not included.
It is found that both the plasma temperature and \emph{EM}
increased significantly from Q to P phase. During the flare phases from P to D2, the temperature stayed roughly constant at
$\sim (5-6) \times 10^{7}$\,K while the \emph{EM} fluctuated by about an order of magnitude around $10^{54}$\,cm$^{-3}$.
Similar evolution trends in both \emph{EM} and temperature were also observed in the X-ray flare of the brown dwarf LP 412-31 \citep{Stelzer06} and          
the X-ray/optical flare of the late M dwarf Proxima Centauri \citep{Reale04}. The relatively large errors of the spectral fitting prohibits more detailed
studies of the evolution structure in the $\log T - \log \sqrt{EM}$ plane, but the general trend is clear. For source distances of 264\,pc (K7) and
100\,pc (M4), the relative positions between data points remains unchanged while the values on the \emph{EM} axis offset by 0.18 and $-$0.24 respectively.

As the current signal-to-noise is not sufficient for detailed modeling, we studied the physical properties of the flare based on the
Haisch's simplified approach (HSA) \citep{Haisch83, CompLoops}. In this simplified parametrized scheme, the length of the flare loop 
can be estimated based only on the $EM$ and the flare decay timescale $\rm \tau_{d}$. 
The loop lengths derived here represent the upper limits since multiple loops could be present while the numbers presented are for the single-loop
scenario; together with the fact that the determined \emph{EM} in the flare phase (F) should also include, albeit small, contributions 
from the precursor phase (P).
In addition to the work by \citet{CompLoops} with \emph{EUVE}, \citet{PandeySingh12} adopted the HSA to study a number of X-ray flares 
from CVn-type binaries captured by \emph{XMM-Newton}, while \citet{Schmitt87} also tested the reliability of HSA with solar X-ray flares 
using \emph{EM} values measured from \emph{Einstein}.
Using the D1 phase decay timescale derived from the MOS lightcurve and the sum of F phase \emph{EM}, the flare loop length during
the F phase is estimated to be $\rm (3.4 - 5.6) \times 10^{10}\,cm$. Taking the stellar radii of M4 and K7 stars to be 0.4\,$\rm R_\odot$
and 0.7\,$\rm R_\odot$ \citep{Lacy77, White89, Segransan03}, the size of the loop is estimated to be 1.1 $-$ 1.3 times of the stellar radius.
The range of the flare loop length in stellar radius is computed by the loop length and the stellar radius estimated for the
associated spectral type.
This is compatible with the picture that long-loop flares on main sequence stars occur mainly on M dwarfs as proposed by \citet{CompLoops}.

\section{Discussion and Conclusions}
\label{Discussions}
In this paper we report the discovery and the multi-wavelength investigation of an X-ray/UV stellar flare from the K7 to M4 dwarf star \mysrc.
The long, continuous X-ray observation of the source from the onset of the flare makes this event unique among those presented in the existing literature.
While concurrent flare observations in X-ray and optical/UV bands had also been reported 
on later type M8 and M8.5 stars \citep{Stelzer06, Robrade10}, the current study is one of the best observed samples of late K and early M dwarfs 
with a full 11-hour X-ray coverage from quiescence to late decay phase together with a well observed precursor. 

Quiescent emission from the source is found from multiple observations in X-ray, optical, and IR,
with a counterpart identified in both the 2MASS and DENIS catalogues. 
The IR photometry and colors characterize the source as an K7 to M4 dwarf star,
supporting the conclusion that the observed high energy flare originated from the stellar coronae.
In quiescence, \mysrc\ has an X-ray luminosity of $L_{\rm X,q} \approx$ (0.2 $-$ 1.7)$\times 10^{29}$\,erg\,s$^{-1}$ 
and a temperature of $\sim$12\,MK from the \emph{XMM-Newton} data. The values are typical
as for M dwarfs \citep{EXOSAT90, Schmitt95, Schmitt04}, and are consistent with those deduced from both \emph{Chandra} and
\emph{Swift} observations in quiescence.
The X-ray luminosity in quiescence and the bolometric luminosity of \mysrc\ are found to be closely consistent with the K7 $-$ M4 subsample studied
by \citet{EXOSAT90} and \citet{Tsikoudi00}.
Moreover, the observed $\log (L_{\rm q}/L_{\rm bol})$ agree almost perfectly with the tight correlation of the two quantities 
deduced by \citet{EXOSAT90} for late-K to M dwarfs, thus further confirming the K7$-$M4 classification of \mysrc.
The distance of \mysrc\ based on spectroscopic parallax and the $n\rm _{H}$ limit was estimated to be 100$-$264\,pc.
Consistent results obtained in the timing and spectral analysis with similar dwarf flares further strengthen the conclusion that \mysrc\ is indeed in 
the foreground instead of residing in the LMC. 

Clear precursor emission before the commencement of the main flare event was observed, 
with released X-ray energy of $\rm \sim(0.5 - 3.2) \times 10^{34}$\,erg, or about one-tenth of the total. 
Similar precursor phases have also been observed from the ultracool dwarf LP 412-31 \citep{Stelzer06} and from the Sun \citep{Joshi11}, 
suggesting an intimate spatial relation between the pre-flare magnetically active regions and the main flare. 
The X-ray luminosity then sharply increased through the R phase to reach values of 
$\rm (1.0 - 7.0) \times 10^{31}$\,$\rm erg\,s^{-1}$ and $\rm (0.3 - 1.7) \times 10^{31}$\,$\rm erg\,s^{-1}$ in X-ray and UV, respectively.
During the flare peak, the X-ray luminosity of \mysrc\ was $\sim$400 times higher than the quiescent value.
The strongest increase recorded by \citet{EXOSAT90} was only $\sim$20, while more recently observed X-ray flares exhibited flux increases by factors
ranging from few \citep{Pandey08} to tens \citep{Schmitt02, Gudel04, Stelzer06, Wargelin08, Robrade10, Gupta11}, 
and up to hundreds \citep{Favata00, Fleming00, Rutledge00} to $\sim$7000 \citep{Osten10}.
Furthermore, the quiescent luminosities of these sources ($10^{25} - 10^{28}$\,$\rm erg\,s^{-1}$)
place \mysrc\ at the energetic end of the X-ray flare samples.
The loop length of \mysrc\ is estimated from the HSA scheme to be comparable or bigger than the stellar radius, 
classifying our current flare under the `long loop' category comparable  
to the $\sim$1$-$3 times stellar radius reported for less luminous flares \citep{Schmitt02,CompLoops, Wargelin08}.

The long-time continuous X-ray coverage of the flare event allows for detailed lightcurve analysis.
After the main flare peak, there could be a brief rise phase, arising possibly from the superposition of a less intense secondary flare on the main one. 
This could imply a more complicated spatial structure of the corona than the compact single-loop flaring scenario.
More prominent multi-peak X-ray/UV flares had been observed for a number of G to K dwarfs \citep{Pandey08} and M dwarfs \citep{GALEXUV}. 
For \mysrc, the drop in flux after the flare peak could be best characterized by a two-component exponential decay with e-folding 
timescales of $\sim$28 min and $\sim$4 hours, and the single-component decay model is ruled out. 
Similar two-stage flare decays were observed in some M dwarf X-ray flares including the 
M6 Proxima Centauri \citep{Reale04}, M3.5 Ross 154 \citep{Wargelin08} and the M8 LP 412$-$31 \citep{Stelzer06}. 
There are different interpretations of the change in decay timescale. Observations of large solar flares suggested that the slower decay may be resulted
from the reconnected loops of the main flare which evolve differently \citep{Aschwanden01} or the overlapping with another ongoing flare event
\citep{Reale04}.
On the other hand, there are flares in which single component decay models provide satisfactory fits. Their decay timescales range from 
$\sim$~0.1\,ks to $\gtrsim$~10\,ks.
Finally, the observed count rate still exceeded the quiescent level by a factor of $\sim$~5 $-$ 7 at the end of observation after nine hours of decay.
Such late-time decay emission could be coming from continuous low level flare events occurring in the active regions left behind 
by the main flare.

The evolution of $n\rm _{H}$ at different phases cannot be constrained with the current dataset.
Spectral fitting with two-temperature APEC model supports the parameterization of the flare spectra into a low-temperature component 
and a time-dependent high-temperature flare component with similar spectral shapes and comparable parameters to other late M objects 
\citep{Hambaryan04, Stelzer06, Robrade09, Robrade10, Gupta11}. 
The cooler components ($\sim$10\,MK) are observed to have enhanced \emph{EM}, along with the emergence of the hot components ($\sim$60\,MK)
with \emph{EM} of up to $(0.4 - 3.1) \times 10^{54}$\,cm$^{-3}$ during flare peak.
The total X-ray energy released is $\sim(0.4 - 2.9) \times 10^{35}$\,erg. While only very few K7$-$M4 dwarf X-ray flares have been reported, 
the observed properties of \mysrc\ place it at the energetic end of the limited sample.

The simultaneous UV observations show significant 
brightening in the UV band. Focusing primarily on the temporal coincidence between lightcurves in X-ray and UV, \citet{MitraKraev05} found an 
approximate time lag of $\sim$10 minutes in X-ray. 
The low sampling rate of our UV data for \mysrc\ could not confirm such possible lag. 
On the other hand, the lower limit of the X-ray to UV energy loss ratio derived for the quiescence phase and the ratio observed during the flare phase 
is consistent with the description of the quisecent state as the superposition of 
continuous low-energy small scale flares known as micro-flares \citep{Parker88}, or nano-flares depending on the timescales.

\clearpage

\begin{deluxetable}{lcccccccc}
\tablewidth{0pt}
\tabletypesize{\tiny}
\tablecaption
{X-ray detection of \mysrc\ by the \emph{XMM-Newton} EPIC
\label{xmmsumtable}
}

\tablehead{
   \colhead{Obs.ID} &
   \colhead{Det-ExpID} &
   \colhead{Date} &
   \colhead{Start Time} &
   \colhead{Exp. Time} &
   \colhead{Mode \tablenotemark{a}} &
   \colhead{Opt. Fil.} &
   \colhead{Det. ML} &
   \colhead{Average Count Rate} \\

   &
   &
   \colhead{(MM/DD/YYYY)} &
   \colhead{(UT HH:MM)} &
   \colhead{(ks)} &
   &
   &
   &
   \colhead{(s$^{-1}$)}

}
\startdata

0113000401 & M2S007 & 05/23/2001 & 07:33 & 47.8 & PFW & Thick & 24.4 & 0.0026 \\ 
0117510201 & PNS010\tablenotemark{b} & 02/11/2000 & 09:05 & 10.8 & PFW & Medium & 55.2 & 0.0172 \\ 
0117730501 & M1S014 & 02/17/2000 & 13:29 & 10.2 & PFW & Medium & 20.8 & 0.0064 \\ 
0117730501 & M1S016 & 02/17/2000 & 17:20 & 14.2 & PPW2 & Medium & 24.1 & 0.0051 \\ 
0125120101 & M1S008 & 05/26/2000 & 12:54 & 31.9 & PFW & Medium & 28.7 & 0.0034 \\ 
0125120101 & M2S010 & 05/26/2000 & 12:54 & 31.9 & PFW & Medium & 27.8 & 0.0035 \\ 
0125120101 & PNS006\tablenotemark{b} & 05/26/2000 & 11:03 & 38.6 & PFW & Medium & 170.6 & 0.0159 \\ 
0125120201 & M1S007\tablenotemark{c} & 05/27/2000 & 01:50 & 40.2 & FastUC & Medium & 6056.0 & 0.1218 \\ 
0125120201 & M2S009\tablenotemark{c} & 05/27/2000 & 01:50 & 40.2 & FastUC & Medium & 6055.7 & 0.1193 \\ 
0413180201 & M2S002 & 10/02/2006 & 01:01 & 13.6 & FastUC & Medium & 9.0 & 0.0027 \\ 
0413180301 & M2S002 & 10/02/2006 & 18:36 & 16.2 & FastUC & Medium & 15.5 & 0.0033 \\

\enddata

\tablenotetext{a}{Science mode of the central CCD. PFW: Prime Full Window; PPW2: Prime Partial W2 (small window);
FastUC: Timing uncompressed. All detections here were made with imaging mode data.}
\tablenotetext{b}{XMM PN Observations capturing the quiescence, used for spectral analysis of quiescence (Q) 
phase (Table \ref{SPECFitTab}).}
\tablenotetext{c}{The only XMM MOS Observations capturing the flare event, used for spectral analysis of the precursor (P),
rise (R), decay 1 and 2 (D1-D2) and flare (F) phases (Table \ref{SPECFitTab}).}

\end{deluxetable}

\begin{deluxetable}{lccccccc}
\tablewidth{0pt}
\tabletypesize{\tiny}
\tablecaption
{UV detection of \mysrc\ by the \emph{XMM-Newton} Optical Monitor
\label{OMUVTab}
}

\tablehead{
   \colhead{OM Exp. ID} &
   \colhead{UV Filter} &
   \colhead{Start Time} &
   \colhead{Exp. Time} &
   \colhead{Count Rate} &
   \colhead{Magnitude} &
   \colhead{Flux} &
   \colhead{Luminosity\tablenotemark{a}} \\
   
   &
   &
   \colhead{(UT HH:MM)} &
   \colhead{(s)} &
   \colhead{(counts/s)} &
   &
   \colhead{($10^{-16}$\,erg\,cm$^{-2}$\,s$^{-1}$\,$\AA^{-1}$)} &
   \colhead{($10^{30}$\,erg\,s$^{-1}$)}

}
\startdata

\multicolumn{8}{c}{0125120101 on 2000 May 26.} \\
440 & UVW2 & 16:40 & 4000 & $\leq 0.07$ & $\geq 17.7$ & $\leq$ 4.2 & $\leq$ 2.3 \\

\multicolumn{8}{c}{0125120201 on 2000 May 27.} \\
432 & UVW1 & 02:18 & 1000 & $\leq 0.38$ & $\geq 18.3$ & $\leq$ 1.8 & $\leq$ 1.4 \\
436\tablenotemark{\dagger} & UVW1 & 04:09 & 1000 & 3.81 $ \pm $ 0.09 & 15.75 $ \pm $ 0.03 & (18.0 $\pm$ 0.5) & (2.5$-$17.4) \\

440 & UVM2 & 07:39 & 2500 & $\leq 0.13$ & $\geq 18.0$ & $\leq$ 2.6 & $\leq$ 1.2 \\
444 & UVW2 & 13:14 & 4000 & $\leq 0.08$ & $\geq 17.6$ & $\leq$ 4.4 & $\leq$ 1.9 \\
448 & UVW2 & 18:44 & 2500 & $\leq 0.09$ & $\geq 17.5$ & $\leq$ 4.7 & $\leq$ 2.0 \\

\multicolumn{8}{c}{011300401 on 2001 May 23.} \\
401 & UVM2 & 08:34 & 3500 & $\leq 0.12$ & $\geq 18.1$ & $\leq$ 2.6 & $\leq$ 1.5 \\
404 & UVM2 & 11:45 & 3500 & $\leq 0.11$ & $\geq 18.2$ & $\leq$ 2.3 & $\leq$ 1.4 \\
405 & UVW2 & 13:53 & 3500 & $\leq 0.08$ & $\geq 17.6$ & $\leq$ 4.6 & $\leq$ 2.5 \\
408 & UVW2 & 17:04 & 3500 & $\leq 0.07$ & $\geq 17.7$ & $\leq$ 4.2 & $\leq$ 2.3 \\

\enddata

\tablecomments{3-$\sigma$ upper limits were computed for non-detections.}
\tablenotetext{a}{The luminosity range is obtained by the assumed distance range of (100$-$264\,pc), while the upper limits
assumed the distance 264\,pc}

\end{deluxetable}

\begin{deluxetable}{lcccccc}
\tablewidth{0pt}
\tabletypesize{\scriptsize}
\tablecaption
{Source positions of \mysrc\ derived from X-ray and NIR instruments
\label{srcpostable}
}

\tablehead{
   \colhead{Instrument} &
   \colhead{Detector} & 
   \colhead{R.A.} &
   \colhead{R.A. Error} &
   \colhead{Dec.} &
   \colhead{Dec. Error} \\
   
   &
   &
   \colhead{(J2000.0)} &
   \colhead{(arcsec)} &
   \colhead{(J2000.0)} &
   \colhead{(arcsec)}

}
\startdata
\multicolumn{6}{c}{X-ray Positions} \\
\hline
\emph{Chandra} & HRC-I & 05 41 45.28 & 1.09 & $-$69 21 51.5 & 1.0 \\ 
\emph{XMM-Newton} (Flare) & MOS & 05 41 45.42 & 1.42 & $-$69 21 49.4 & 1.4 \\ 
\emph{XMM-Newton} (Quiescent) & MOS + PN & 05 41 45.29 & 0.86 & $-$69 21 50.6 & 0.9 \\ 
\emph{Swfit} & XRT-PC & 05 41 45.35 & 3.60 & $-$69 21 51.2 & 3.6 \\ 
ROSAT & HRI & 05 41 44.60 & 4.40 & $-$69 21 51.1 & 4.4 \\

\hline
\hline
\multicolumn{6}{c}{NIR Positions} \\
\hline
\multicolumn{2}{c}{2MASS} & 05 41 45.34 & 0.19 & $-$69 21 51.22 & 0.1 \\ 
\multicolumn{2}{c}{DENIS} & 05 41 45.27 & $\le 1$ & $-$69 21 51.10 & $\le 1 $ \\

\enddata
\tablecomments{NIR counterparts are 2MASS J05414534$-$6921512 and DENIS J054145.2$-$692151.}

\end{deluxetable}

\begin{deluxetable}{lccc}
\tablewidth{0pt}
\tabletypesize{\scriptsize}
\tablecaption
{NIR magnitudes and colors of the 2MASS and DENIS counterparts
\label{counterpartinfo}
}

\tablehead{
   \colhead{2MASS Counterpart} &
   \colhead{$J$} &
   \colhead{$H$} &
   \colhead{$K_{\rm S} $} \\
   
   &
   \colhead{$J - H$} &
   \colhead{$H - K_{\rm S} $} &
   \colhead{$J - K_{\rm S} $} 

}
\startdata

2MASS J05414534$-$6921512 & 13.31 $ \pm $ 0.04 & 12.68 $ \pm $ 0.05 & 12.47 $ \pm $ 0.04 \\
                          & 0.63 $ \pm $ 0.06 & 0.21 $ \pm $ 0.06 & 0.84 $ \pm $ 0.06 \\

\hline
DENIS Counterpart & $I$ & $J$ & $K_{\rm S}$ \\
 & $I - J$ & $J - K_{\rm S}$ & $I - K_{\rm S}$ \\
\hline

DENIS J054145.2$-$692151 (a) & 14.38 $ \pm $ 0.03 & 13.19 $ \pm $ 0.08 & 12.41 $ \pm $ 0.14 \\
 & 1.20 $ \pm $ 0.09 & 0.78 $ \pm $ 0.16 & 1.98 $ \pm $ 0.14\\
DENIS J054145.2$-$692151 (b) & 14.34 $ \pm $ 0.03 & 13.20 $ \pm $ 0.07 & 12.16 $ \pm $ 0.11 \\
 & 1.14 $ \pm $ 0.08 & 1.04 $ \pm $ 0.13 & 2.18 $ \pm $ 0.11\\

\enddata

\end{deluxetable}

\begin{deluxetable}{lccccccc}
\tablewidth{0pt}
\tabletypesize{\tiny}
\tablecaption
{Spectral Fitting Results of \mysrc's flare with the \emph{XMM-Newton} detections
\label{SPECFitTab}
}

\tablehead{
   \colhead{Stage} & 
   \colhead{Duration $t_{0} +$} &
   \colhead{n$\rm _{H}$} & 
   \colhead{kT} & 
   \colhead{$\chi ^{2}_{\nu}$ (D.O.F.)} &
   \colhead{EM\tablenotemark{c}} &
   \colhead{Flux\tablenotemark{d}} &
   \colhead{Luminosity\tablenotemark{c,d}} \\
   
   &
   \colhead{(Hours)} &
   \colhead{(10$^{20}$\,cm\,$^{-1}$)} &
   \colhead{(keV)} &
   &
   \colhead{($10^{52}$\,cm$^{-3}$)} &
   \colhead{($10^{-13}$\,erg\,cm$^{-2}$\,s$^{-1}$)} &
   \colhead{($10^{29}$\,erg\,s$^{-1}$)} \\

}
\startdata

\multicolumn{8}{c}{1T APEC Model Results} \\
Q & \dotfill \tablenotemark{a} & 0.67\tablenotemark{b} & 1.22$^{+0.11}_{-0.23} $ & 1.36 (22) & (0.1 $-$ 0.7)  & 
(0.21 $\pm$ 0.03) & (0.2 $-$ 1.7) \\

P & 0.00 $-$ 0.90 &  0.67\tablenotemark{b} & 4.33$^{+2.27}_{-1.13} $ & 1.61 (10) & (6.7 $-$ 46.5)  & 
(11.2 $\pm$ 1.8)  & (13.3 $-$ 92.9) \\

R &  0.90 $-$ 1.13 &  2.09$^{+2.13}_{-1.45} $ & 4.88$^{+1.88}_{-1.14} $ & 0.99 (17) & (47.4 $-$ 330) & 
(82.0 $\pm$ 9.8)  & (97.7 $-$ 681) \\

D1 & 1.13 $-$ 1.59  & 0.71 $^{+0.79}_{-0.60} $  & 3.95 $^{+0.62}_{-0.55} $ & 1.44 (37) & (42.4 $-$ 295) & 
(70.0 $\pm$ 5.5)  & (83.4 $-$ 581) \\

D2 & 1.59 $-$ 10.10  & 0.57 $^{+0.50}_{-0.40} $  & 3.29$^{+0.30}_{-0.27} $ & 1.94 (60) & (4.1 $-$ 28.8) & 
(6.6 $\pm$ 0.4)  & (7.8 $-$ 54.6) \\

F & 0.96 $-$ 1.52  & 1.23$^{+0.75}_{-0.60} $ & 4.29$^{+0.71}_{-0.48} $ & 1.19 (49) & (49.5 $-$ 345) & 
(83.4 $\pm$ 5.6) & (99.4 $-$ 693) \\

\multicolumn{8}{c}{2T APEC Model Results} \\
P & 0.00 $-$ 0.90 & 0.67\tablenotemark{b} & 0.28 $^{+1.20}_{-0.28}$ & 1.45 (8) & (0.4 $-$ 2.8) & 
$(11.7 \pm 2.1)$ & (14.0 $-$ 97.4) \\
  & & & 5.21 $^{+2.47}_{-1.82}$ & & (6.1 $-$ 42.6) & & \\

R & 0.90 $-$ 1.13 & 1.91 $^{+1.92}_{-1.35}$& 1.01 $^{+3.32}_{-0.69}$ & 0.98 (15) & (1.5 $-$ 10.2) & 
$(83.2 \pm 10.1)$ & (99.5 $-$ 693) \\
  & & & 5.83 $^{+2.31}_{-1.25}$ & & (44.2 $-$ 308) & & \\

D1 & 1.13 $-$ 1.59 & 0.67$^{+0.88}_{-0.63}$ & 0.95 $^{+0.13}_{-0.17}$ & 0.93 (35) & (2.6 $-$ 18.2) & 
$(70.8 \pm 5.8)$ & (84.7 $-$ 591) \\
  & & & 4.66 $^{+1.39}_{-0.80}$ & & (36.9 $-$ 257) & & \\

D2 & 1.59 $-$ 10.10 & 0.67$^{+0.57}_{-0.45}$ & 0.76 $^{+0.10}_{-0.16}$ & 1.09 (58) & (0.3 $-$ 1.8) & 
$(6.7 \pm 0.40)$ & (8.0 $-$ 55.6) \\
  & & & 3.61 $^{+0.49}_{-0.36}$ & & (3.6 $-$ 25.1) & & \\

F & 0.96 $-$ 1.52 & 1.14$^{+0.80}_{-0.63}$ & 1.06 $^{+0.28}_{-0.22}$ & 0.99 (47) & (2.3 $-$ 16.1) & 
$(84.2 \pm 5.9)$ & (101 $-$ 702) \\
  & & & 5.07 $^{+1.61}_{-0.86}$ & & (44.8 $-$ 312) & & \\

\enddata

\tablecomments{The error ranges of all parameters give the 90\% confidence intervals.}
\tablenotetext{a}{The spectral properties are derived from previous non-flare PN observations, as listed in Table \ref{xmmsumtable}.}
\tablenotetext{b}{The value was fixed to the average of the Decay stages' in the 2T model.}
\tablenotetext{c}{Parameter are shown in ranges correponding to source distance of 100$-$264\,pc.}
\tablenotetext{d}{X-ray fluxes and luminosities are for the 0.1 $-$ 7.0 keV band.}

\end{deluxetable}

\acknowledgments
We would like to thank the anonymous referee for the helpful comments on the characterization of the spectral type of the source.
B. T. H. Tsang acknowledges valuable discussions with Dr. Jifeng Liu and Dr. Frank Primini, and the support from HKU
under the grant of Postgraduate Studentship.
A. K. H. Kong gratefully acknowledges support from the National Science Council of the Republic of China (Taiwan) through grant
NSC100-2628-M-007-002-MY3 and a Kenda Foundation Golden Jade Fellowship.
This research has made use of data and software provided by the High Energy Astrophysics Science Archive Research Center (HEASARC),
which is a service of the Astrophysics Science Division at NASA/GSFC and the High Energy Astrophysics Division of the Smithsonian Astrophysical Observatory.
This research has also made use of the NASA/IPAC Infrared Science Archive, which is operated by the Jet Propulsion Laboratory,
California Institute of Technology, under contract with the National Aeronautics and Space Administration.
This was also supported by the use of observations made with the European Southern Observatory telescopes
obtained from the ESO/ST-ECF Science Archive Facility.
This investigation was supported by the RGC/GRF grant 704709 from the government of the Hong Kong SAR.

\markboth{Bibliography}{Bibliography}
\bibliography{paper}

\begin{thebibliography}{57}
\expandafter\ifx\csname natexlab\endcsname\relax\def\natexlab#1{#1}\fi

\bibitem[{{Allred} {et~al.}(2006){Allred}, {Hawley}, {Abbett}, \&
  {Carlsson}}]{Allred06}
{Allred}, J.~C., {Hawley}, S.~L., {Abbett}, W.~P., \& {Carlsson}, M. 2006,
  \apj, 644, 484

\bibitem[{{Aschwanden} \& {Alexander}(2001)}]{Aschwanden01}
{Aschwanden}, M.~J., \& {Alexander}, D. 2001, \solphys, 204, 91

\bibitem[{{Audard} {et~al.}(2003){Audard}, {G{\"u}del}, {Sres}, {Raassen}, \&
  {Mewe}}]{Audard03}
{Audard}, M., {G{\"u}del}, M., {Sres}, A., {Raassen}, A.~J.~J., \& {Mewe}, R.
  2003, \aap, 398, 1137

\bibitem[{{Benz} \& {G{\"u}del}(2010)}]{Benz10}
{Benz}, A.~O., \& {G{\"u}del}, M. 2010, \araa, 48, 241

\bibitem[{{Berger} {et~al.}(2010){Berger}, {Basri}, {Fleming}, {Giampapa},
  {Gizis}, {Liebert}, {Mart{\'{\i}}n}, {Phan-Bao}, \& {Rutledge}}]{Berger10}
{Berger}, E., {et~al.} 2010, \apj, 709, 332

\bibitem[{{Cabrera-Lavers} \& {Garz{\'o}n}(2003)}]{Cabrera03}
{Cabrera-Lavers}, A., \& {Garz{\'o}n}, F. 2003, \aap, 403, 383

\bibitem[{{DENIS Consortium}(2005)}]{DENIS}
{DENIS Consortium}. 2005, VizieR Online Data Catalog, 2263, 0

\bibitem[{{Dickey} \& {Lockman}(1990)}]{DLMap}
{Dickey}, J.~M., \& {Lockman}, F.~J. 1990, \araa, 28, 215

\bibitem[{{Favata} {et~al.}(2001){Favata}, {Micela}, \& {Reale}}]{Favata01}
{Favata}, F., {Micela}, G., \& {Reale}, F. 2001, \aap, 375, 485

\bibitem[{{Favata} {et~al.}(2000){Favata}, {Reale}, {Micela}, {Sciortino},
  {Maggio}, \& {Matsumoto}}]{Favata00}
{Favata}, F., {Reale}, F., {Micela}, G., {Sciortino}, S., {Maggio}, A., \&
  {Matsumoto}, H. 2000, \aap, 353, 987

\bibitem[{{Fleming} {et~al.}(2000){Fleming}, {Giampapa}, \&
  {Schmitt}}]{Fleming00}
{Fleming}, T.~A., {Giampapa}, M.~S., \& {Schmitt}, J.~H.~M.~M. 2000, \apj, 533,
  372

\bibitem[{{Gelino} {et~al.}(2009){Gelino}, {Kirkpatrick}, \&
  {Burgasser}}]{DwarfDatabase}
{Gelino}, C.~R., {Kirkpatrick}, J.~D., \& {Burgasser}, A.~J. 2009, in American
  Institute of Physics Conference Series, Vol. 1094, American Institute of
  Physics Conference Series, ed. {E.~Stempels}, 924--927

\bibitem[{{Guainazzi}(2011)}]{Guainazzi11}
{Guainazzi}, M. 2011, XMM-Newton Calibration Technical Note, Tech. Rep.
  XMM-SOC-CAL-TN-0018, ESA-ESAC, Villafranca del Castillo, Spain

\bibitem[{{G{\"u}del} {et~al.}(2004){G{\"u}del}, {Audard}, {Reale}, {Skinner},
  \& {Linsky}}]{Gudel04}
{G{\"u}del}, M., {Audard}, M., {Reale}, F., {Skinner}, S.~L., \& {Linsky},
  J.~L. 2004, \aap, 416, 713

\bibitem[{{Gupta} {et~al.}(2011){Gupta}, {Galeazzi}, \& {Williams}}]{Gupta11}
{Gupta}, A., {Galeazzi}, M., \& {Williams}, B. 2011, \apj, 731, 63

\bibitem[{{Haisch}(1983)}]{Haisch83}
{Haisch}, B.~M. 1983, in IAU Colloq. 71, Activity in Red Dwarf Stars, ed.
  {{Byrne}, P. B., {Rodono}, M.}, 255

\bibitem[{{Hambaryan} {et~al.}(2004){Hambaryan}, {Staude}, {Schwope}, {Scholz},
  {Kimeswenger}, \& {Neuh{\"a}user}}]{Hambaryan04}
{Hambaryan}, V., {Staude}, A., {Schwope}, A.~D., {Scholz}, R.-D.,
  {Kimeswenger}, S., \& {Neuh{\"a}user}, R. 2004, \aap, 415, 265

\bibitem[{{Hawley} {et~al.}(2002){Hawley}, {Covey}, {Knapp}, {Golimowski},
  {Fan}, {Anderson}, {Gunn}, {Harris}, {Ivezi{\'c}}, {Long}, {Lupton},
  {McGehee}, {Narayanan}, {Peng}, {Schlegel}, {Schneider}, {Spahn}, {Strauss},
  {Szkody}, {Tsvetanov}, {Walkowicz}, {Brinkmann}, {Harvanek}, {Hennessy},
  {Kleinman}, {Krzesinski}, {Long}, {Neilsen}, {Newman}, {Nitta}, {Snedden}, \&
  {York}}]{MLTSDSS}
{Hawley}, S.~L., {et~al.} 2002, \aj, 123, 3409

\bibitem[{{Huenemoerder} {et~al.}(2001){Huenemoerder}, {Canizares}, \&
  {Schulz}}]{Huenemoerder01}
{Huenemoerder}, D.~P., {Canizares}, C.~R., \& {Schulz}, N.~S. 2001, \apj, 559,
  1135

\bibitem[{{Joshi} {et~al.}(2011){Joshi}, {Veronig}, {Lee}, {Bong}, {Tiwari}, \&
  {Cho}}]{Joshi11}
{Joshi}, B., {Veronig}, A.~M., {Lee}, J., {Bong}, S.-C., {Tiwari}, S.~K., \&
  {Cho}, K.-S. 2011, \apj, 743, 195

\bibitem[{{Kahn} {et~al.}(1979){Kahn}, {Mason}, {Bowyer}, {Linsky}, {Haisch},
  {White}, \& {Pravdo}}]{HEAO1det}
{Kahn}, S.~M., {Mason}, K.~O., {Bowyer}, C.~S., {Linsky}, J.~L., {Haisch},
  B.~M., {White}, N.~E., \& {Pravdo}, S.~H. 1979, \apj, 234, 107

\bibitem[{{Kalberla} {et~al.}(2005){Kalberla}, {Burton}, {Hartmann}, {Arnal},
  {Bajaja}, {Morras}, \& {P{\"o}ppel}}]{LABSur}
{Kalberla}, P.~M.~W., {Burton}, W.~B., {Hartmann}, D., {Arnal}, E.~M.,
  {Bajaja}, E., {Morras}, R., \& {P{\"o}ppel}, W.~G.~L. 2005, \aap, 440, 775

\bibitem[{{Lacy}(1977)}]{Lacy77}
{Lacy}, C.~H. 1977, \apjs, 34, 479

\bibitem[{{Leggett} {et~al.}(2001){Leggett}, {Allard}, {Geballe}, {Hauschildt},
  \& {Schweitzer}}]{BCKdML}
{Leggett}, S.~K., {Allard}, F., {Geballe}, T.~R., {Hauschildt}, P.~H., \&
  {Schweitzer}, A. 2001, \apj, 548, 908

\bibitem[{{Mitra-Kraev} {et~al.}(2005){Mitra-Kraev}, {Harra}, {G{\"u}del},
  {Audard}, {Branduardi-Raymont}, {Kay}, {Mewe}, {Raassen}, \& {van
  Driel-Gesztelyi}}]{MitraKraev05}
{Mitra-Kraev}, U., {et~al.} 2005, \aap, 431, 679

\bibitem[{{Mullan} {et~al.}(2006){Mullan}, {Mathioudakis}, {Bloomfield}, \&
  {Christian}}]{CompLoops}
{Mullan}, D.~J., {Mathioudakis}, M., {Bloomfield}, D.~S., \& {Christian}, D.~J.
  2006, \apjs, 164, 173

\bibitem[{{Nikolaev} \& {Weinberg}(2000)}]{stellarpoplmctmass}
{Nikolaev}, S., \& {Weinberg}, M.~D. 2000, \apj, 542, 804

\bibitem[{{Osten} {et~al.}(2006){Osten}, {Hawley}, {Allred}, {Johns-Krull},
  {Brown}, \& {Harper}}]{Osten06}
{Osten}, R.~A., {Hawley}, S.~L., {Allred}, J., {Johns-Krull}, C.~M., {Brown},
  A., \& {Harper}, G.~M. 2006, \apj, 647, 1349

\bibitem[{{Osten} {et~al.}(2010){Osten}, {Godet}, {Drake}, {Tueller},
  {Cummings}, {Krimm}, {Pye}, {Pal'shin}, {Golenetskii}, {Reale}, {Oates},
  {Page}, \& {Melandri}}]{Osten10}
{Osten}, R.~A., {et~al.} 2010, \apj, 721, 785

\bibitem[{{Pallavicini} {et~al.}(1990){Pallavicini}, {Tagliaferri}, \&
  {Stella}}]{EXOSAT90}
{Pallavicini}, R., {Tagliaferri}, G., \& {Stella}, L. 1990, \aap, 228, 403

\bibitem[{{Pandey} \& {Singh}(2008)}]{Pandey08}
{Pandey}, J.~C., \& {Singh}, K.~P. 2008, \mnras, 387, 1627

\bibitem[{{Pandey} \& {Singh}(2012)}]{PandeySingh12}
---. 2012, \mnras, 419, 1219

\bibitem[{{Parker}(1988)}]{Parker88}
{Parker}, E.~N. 1988, \apj, 330, 474

\bibitem[{{Priest} \& {Forbes}(2002)}]{Priest02}
{Priest}, E.~R., \& {Forbes}, T.~G. 2002, \aapr, 10, 313

\bibitem[{{Raassen} {et~al.}(2003){Raassen}, {Mewe}, {Audard}, \&
  {G{\"u}del}}]{Raassen03}
{Raassen}, A.~J.~J., {Mewe}, R., {Audard}, M., \& {G{\"u}del}, M. 2003, \aap,
  411, 509

\bibitem[{{Reale} {et~al.}(1997){Reale}, {Betta}, {Peres}, {Serio}, \&
  {McTiernan}}]{Reale97}
{Reale}, F., {Betta}, R., {Peres}, G., {Serio}, S., \& {McTiernan}, J. 1997,
  \aap, 325, 782

\bibitem[{{Reale} {et~al.}(2004){Reale}, {G{\"u}del}, {Peres}, \&
  {Audard}}]{Reale04}
{Reale}, F., {G{\"u}del}, M., {Peres}, G., \& {Audard}, M. 2004, \aap, 416, 733

\bibitem[{{Robrade} {et~al.}(2010){Robrade}, {Poppenhaeger}, \&
  {Schmitt}}]{Robrade10}
{Robrade}, J., {Poppenhaeger}, K., \& {Schmitt}, J.~H.~M.~M. 2010, \aap, 513,
  A12

\bibitem[{{Robrade} \& {Schmitt}(2009)}]{Robrade09}
{Robrade}, J., \& {Schmitt}, J.~H.~M.~M. 2009, \aap, 496, 229

\bibitem[{{Rutledge} {et~al.}(2000){Rutledge}, {Basri}, {Mart{\'{\i}}n}, \&
  {Bildsten}}]{Rutledge00}
{Rutledge}, R.~E., {Basri}, G., {Mart{\'{\i}}n}, E.~L., \& {Bildsten}, L. 2000,
  \apjl, 538, L141

\bibitem[{{Sanz-Forcada} {et~al.}(2004){Sanz-Forcada}, {Favata}, \&
  {Micela}}]{Sanz-Forcada04}
{Sanz-Forcada}, J., {Favata}, F., \& {Micela}, G. 2004, \aap, 416, 281

\bibitem[{{Sasaki} {et~al.}(2000){Sasaki}, {Haberl}, \& {Pietsch}}]{Sasaki00}
{Sasaki}, M., {Haberl}, F., \& {Pietsch}, W. 2000, \aaps, 143, 391

\bibitem[{{Schmitt} {et~al.}(1987){Schmitt}, {Fink}, \& {Harnden}}]{Schmitt87}
{Schmitt}, J.~H.~M.~M., {Fink}, H., \& {Harnden}, Jr., F.~R. 1987, \apj, 322,
  1023

\bibitem[{{Schmitt} {et~al.}(1995){Schmitt}, {Fleming}, \&
  {Giampapa}}]{Schmitt95}
{Schmitt}, J.~H.~M.~M., {Fleming}, T.~A., \& {Giampapa}, M.~S. 1995, \apj, 450,
  392

\bibitem[{{Schmitt} \& {Liefke}(2002)}]{Schmitt02}
{Schmitt}, J.~H.~M.~M., \& {Liefke}, C. 2002, \aap, 382, L9

\bibitem[{{Schmitt} \& {Liefke}(2004)}]{Schmitt04}
---. 2004, \aap, 417, 651

\bibitem[{{S{\'e}gransan} {et~al.}(2003){S{\'e}gransan}, {Kervella},
  {Forveille}, \& {Queloz}}]{Segransan03}
{S{\'e}gransan}, D., {Kervella}, P., {Forveille}, T., \& {Queloz}, D. 2003,
  \aap, 397, L5

\bibitem[{{Skrutskie} {et~al.}(2006){Skrutskie}, {Cutri}, {Stiening},
  {Weinberg}, {Schneider}, {Carpenter}, {Beichman}, {Capps}, {Chester},
  {Elias}, {Huchra}, {Liebert}, {Lonsdale}, {Monet}, {Price}, {Seitzer},
  {Jarrett}, {Kirkpatrick}, {Gizis}, {Howard}, {Evans}, {Fowler}, {Fullmer},
  {Hurt}, {Light}, {Kopan}, {Marsh}, {McCallon}, {Tam}, {Van Dyk}, \&
  {Wheelock}}]{tMASS}
{Skrutskie}, M.~F., {et~al.} 2006, \aj, 131, 1163

\bibitem[{{Stelzer} {et~al.}(2006){Stelzer}, {Schmitt}, {Micela}, \&
  {Liefke}}]{Stelzer06}
{Stelzer}, B., {Schmitt}, J.~H.~M.~M., {Micela}, G., \& {Liefke}, C. 2006,
  \aap, 460, L35

\bibitem[{{Trenholme} {et~al.}(2004){Trenholme}, {Ramsay}, \&
  {Foley}}]{Treholme04}
{Trenholme}, D., {Ramsay}, G., \& {Foley}, C. 2004, \mnras, 355, 1125

\bibitem[{{Tsikoudi} {et~al.}(2000){Tsikoudi}, {Kellett}, \&
  {Schmitt}}]{Tsikoudi00}
{Tsikoudi}, V., {Kellett}, B.~J., \& {Schmitt}, J.~H.~M.~M. 2000, \mnras, 319,
  1136

\bibitem[{{Tsuboi} {et~al.}(2000){Tsuboi}, {Imanishi}, {Koyama}, {Grosso}, \&
  {Montmerle}}]{Tsuboi00}
{Tsuboi}, Y., {Imanishi}, K., {Koyama}, K., {Grosso}, N., \& {Montmerle}, T.
  2000, \apj, 532, 1089

\bibitem[{{van den Oord} \& {Mewe}(1989)}]{vandenOord89}
{van den Oord}, G.~H.~J., \& {Mewe}, R. 1989, \aap, 213, 245

\bibitem[{{Wargelin} {et~al.}(2008){Wargelin}, {Kashyap}, {Drake},
  {Garc{\'{\i}}a-Alvarez}, \& {Ratzlaff}}]{Wargelin08}
{Wargelin}, B.~J., {Kashyap}, V.~L., {Drake}, J.~J., {Garc{\'{\i}}a-Alvarez},
  D., \& {Ratzlaff}, P.~W. 2008, \apj, 676, 610

\bibitem[{{Watson} {et~al.}(2009){Watson}, {Schr{\"o}der}, {Fyfe}, {Page},
  {Lamer}, {Mateos}, {Pye}, {Sakano}, {Rosen}, {Ballet}, {Barcons}, {Barret},
  {Boller}, {Brunner}, {Brusa}, {Caccianiga}, {Carrera}, {Ceballos}, {Della
  Ceca}, {Denby}, {Denkinson}, {Dupuy}, {Farrell}, {Fraschetti}, {Freyberg},
  {Guillout}, {Hambaryan}, {Maccacaro}, {Mathiesen}, {McMahon}, {Michel},
  {Motch}, {Osborne}, {Page}, {Pakull}, {Pietsch}, {Saxton}, {Schwope},
  {Severgnini}, {Simpson}, {Sironi}, {Stewart}, {Stewart}, {Stobbart}, {Tedds},
  {Warwick}, {Webb}, {West}, {Worrall}, \& {Yuan}}]{tXMM}
{Watson}, M.~G., {et~al.} 2009, \aap, 493, 339

\bibitem[{{Welsh} {et~al.}(2007){Welsh}, {Wheatley}, {Seibert}, {Browne},
  {West}, {Siegmund}, {Barlow}, {Forster}, {Friedman}, {Martin}, {Morrissey},
  {Small}, {Wyder}, {Schiminovich}, {Neff}, \& {Rich}}]{GALEXUV}
{Welsh}, B.~Y., {et~al.} 2007, \apjs, 173, 673

\bibitem[{{White} {et~al.}(1989){White}, {Jackson}, \& {Kundu}}]{White89}
{White}, S.~M., {Jackson}, P.~D., \& {Kundu}, M.~R. 1989, \apjs, 71, 895

\end{thebibliography}

\end{document}